\documentclass[fleqn,usenatbib]{mnras}

\usepackage{newtxtext,newtxmath}

\usepackage[T1]{fontenc}

\DeclareRobustCommand{\VAN}[3]{#2}
\let\VANthebibliography\thebibliography
\def\thebibliography{\DeclareRobustCommand{\VAN}[3]{##3}\VANthebibliography}

\usepackage{graphicx}	
\usepackage{amsmath}	

\title[Cooling of a coronal null]{Heating and cooling at a coronal magnetic null}

\author[P.J. ~Cargill et al.]{
P.~J. Cargill,$^{1,2}$\thanks{E-mail: pcargill@st-andrews.ac.uk}
A.~W. Hood,$^{1}$
D.~Johnson,$^{1}$
\\
$^{1}$School of Mathematics and Statistics, University of St Andrews, St Andrews, Fife, KY16 9SS, UK\\
$^{2}$The Blackett Laboratory, Imperial College, London SW7 2BW, UK
}

\date{Accepted XXX. Received YYY; in original form ZZZ}

\pubyear{2025}

\begin{document}
\label{firstpage}
\pagerange{\pageref{firstpage}--\pageref{lastpage}}
\maketitle

\begin{abstract}
Conductive cooling of the solar corona at a magnetic null is examined. An initial equilibrium is set up, balancing thermal conduction and a constant, spatially uniform coronal heating. The heating is then turned off and the subsequent conductive cooling calculated. An equation for the cooling is obtained using the method of separation of variables and it is shown that the equations for the equilibrium between conduction and heating, and the time-dependent cooling are mathematically identical with a simple change of variables. Thus the properties of the cooling phase are automatically determined by the equilibrium state. For a two-dimensional null, the characteristic cooling timescale increases over that in a straight field by a factor of between 2 and 5, with a scaling determined by the ratio of the {\it{average}} and base areas of a flux element. There is {\it{no explicit dependence}} on the very large areas that can arise near the null. 

\end{abstract}




\section{Introduction} \label{sec:Introduction}

In the solar corona, magnetic nulls are locations where the magnetic field vanishes. They can arise in two- or three-dimensional geometries \citep[e.g.][]{priestforbes} and form an important part of the coronal magnetic field structure. Magnetic fields at the photosphere appear as discrete sources arising as positive and negative polarity pairs. Four sources of similar flux will have at least one null point. In addition, such null points are coronal regions where heating is likely to occur. Indeed, recent work by \citet{cheng2023} identified a local \lq hot\rq \  spot at a coronal null point with sustained emission around 10 MK seen from the AIA 131 {\AA} channel. 

In general terms, the plasma properties of a coronal null are determined by the energy balance between localised plasma heating, and cooling by thermal conduction and optically thin radiation. A first approach was due to \citet{johnson_2024}, who derived analytic solutions for the equilibrium temperature in the vicinity of a two-dimensional null by balancing conductive losses and two different coronal heating functions (both constant in time but one spatially uniform and one with magnitude proportional to $B^2$). The temperature profiles arising from each heating function differ considerably, and include singular solutions in the former case that are removed
by perpendicular thermal conductivity. 

This paper develops the work of \citet{johnson_2024} and considers the conductive cooling of plasma around a two-dimensional null. The null magnetic field profile makes conductive cooling there especially interesting because the field strength varies rapidly. In turn, this implies that the cross-sectional area of an elemental flux tube associated with a given field line also varies. It is well known that conductive cooling of a hot plasma can be strongly dependent on this cross-sectional area, with an increase in area in the region of hot plasma inhibiting conductive cooling (e.g. \cite{as_76}; \cite{cargill_2022}). In turn this raises the possibility of the lifetime of hot plasma near a null being extended from that associated with a straight field as in, for example, a braiding model of coronal heating, and may be of importance in accounting for the faint active region emission from hot plasma in excess of 5 MK \citep[e.g.][]{reale_2009,testa_2011,schmelz_2012}. Accounting for this emission in a braided field has some difficulties due to the very effective field-aligned conduction in a roughly uniform magnetic field (see discussion in \citet{barnes_2016a}). Hence, alternative field geometries might provide a longer-lived hot plasma.

Conductive cooling around a null is investigated by taking the two equilibria of \citet{johnson_2024} and studying the cooling after the heating is turned off. As before, optically thin radiation is neglected. We use a conductive cooling model originally proposed by \citet{as_76}  hereafter AS76. However, for the case of spatially uniform heating, there is an interesting (and quite general) \lq short cut\rq{ } for solving conductive cooling that does not require a full solution of the cooling equation. Section 2 outlines the magnetic field and cooling models used, including the addition of a \lq guide field\rq{ }to the work of \citet{johnson_2024}. Section 3 presents the results for all cases, and shows that simple scalings can account for some, though not all, of the results. Section 4 discusses the distribution of hot plasma around the null during the cooling phase and Section 5 summarises some limitations of the model.

\section{Two-dimensional null equilibria}

\begin{figure*}

  \includegraphics[width=0.4\linewidth]{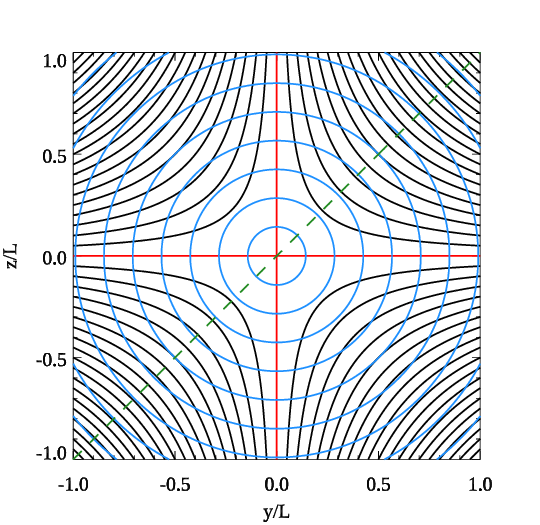}
   \includegraphics[width=0.5\linewidth]{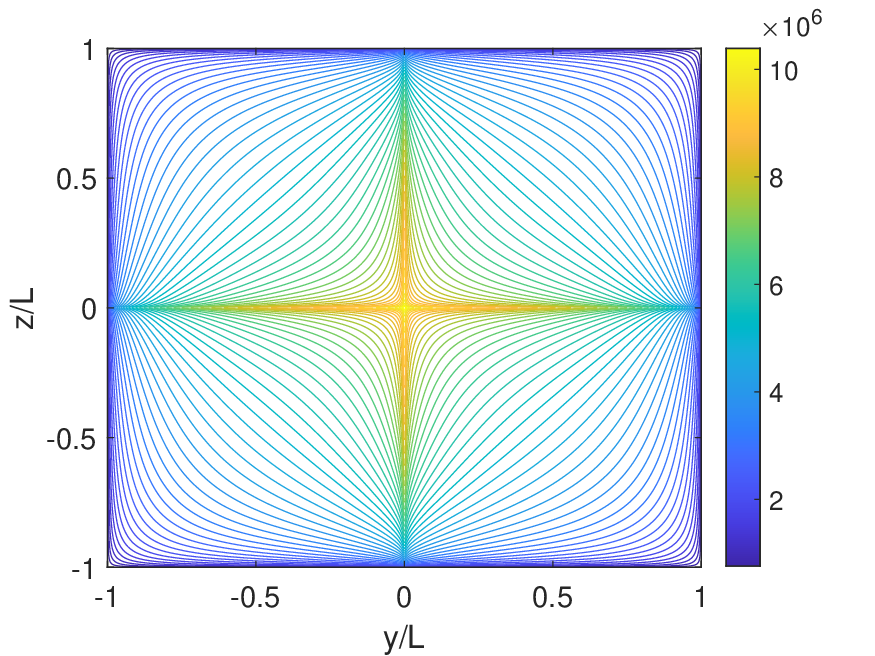}
   
	\caption{Left column: The magnetic field lines (black) and field strength (blue) in the neighbourhood of a magnetic null, located at $y = z = 0$. The dashed line corresponds to the symmetry point of each field line. Right column: a contour plot of the equilibrium temperature for the case with constant heating. In each quadrant the solutions for the temperature are symmetric about the line $|y| = |z|$. In this paper we focus on the upper right quadrant such that y and z are positive. The maximum temperature is of order 10 MK. Brighter colours are associated with higher temperatures.}
	\label{fig:null_sketch}
\end{figure*}

\citet{johnson_2024} obtained static plasma equilibria between a heating term and conductive cooling in a magnetic field with a null point. 
In this paper, we will be interested in plasma with temperature of order 10 MK, density of $10^9 \rm cm^{-3}$ and characteristic length scales of 25 Mm. For these parameters, optically thin radiation can be neglected \citep[e.g.][]{cargill_1993}. Indeed in a plasma heated to a high enough temperature, the radiative cooling phase will always follow the conductive cooling phase \citep{cargill_1995}. The equilibria are then given by:
\begin{equation}
\label{eq:en1}
{\bf B}\cdot \nabla \left( \frac{\kappa_0 T^{5/2}}{B^2} {\bf B} \cdot \nabla T\right) = - H_0 f(B)\; ,
\end{equation}
where $T$ is the temperature, ${\bf B}$ and B the field vector and its magnitude respectively, $\kappa_0 = 8 \times 10^{-7}$ in cgs units \citep[e.g.][]{sb_pc2013} and $H_0 f(B)$ is the heating per unit volume with $f(B)$ permitting a dependence on $B$. 
Normalising the spatial dimensions to a length $L$, which is taken as the half-width of the model box (see \autoref{fig:null_sketch}), and simplifying the equations by defining the quantity $G = T^{7/2}$ gives:
\begin{equation}
\label{eq:en2}
{\bf B}\cdot \nabla \left( \frac{1}{B^2} {\bf B} \cdot \nabla G\right) = - \frac{7H_0L^2}{2 \kappa_0} f(B) = - H^* f(B)\; ,
\end{equation}
where $H^* = 7H_0L^2/2\kappa_0$ is the heating in units of $G$.

The magnetic field around a two-dimensional magnetic null is defined in cartesian coordinates, ${\bf r} = (x, y, z)$, as:
\begin{equation}
\label{eq:B1}
    {{\bf B} = \frac{B_0}{L} (0, y, -z)}\; ,
\end{equation}
where $y$ and $z$ lie between -1 and 1 and $B_0/L$ is a constant whose value is arbitrary since it appears in both the numerator and denominator in \eqref{eq:en1}. For simplicity we set it to unity. This field can also be defined by the x-component of the vector potential: $A_x = yz$\footnote{The symbol $A$ is retained for the cross-sectional area of a flux element}. The field lines and field strength are shown in the left panel of \autoref{fig:null_sketch}. For convenience in this paper, we consider only the upper right quadrant in this figure with $y$ and $z$ positive. A field line is defined as starting on the upper boundary at a point $y = y_0, z = 1, A_x = y_0$, where $0 < y_0 < 1$, and reaches the right hand side boundary at the point $y = 1, z = y_0, A_x = y_0$. The mid-point of the field line is at $y = z = \sqrt{y_0}$, and the temperature profile is symmetric about this point. Hereafter, we refer to this as the symmetry point of any field line. Thus, one need only solve for the temperature between the upper boundary and the symmetry point. [There is a direct analogy with the temperature distribution along magnetic field lines in closed coronal loop models where the temperature is fixed at the photospheric boundaries and the temperature gradient is zero at the loop apex.]

For the null point magnetic field, the length of a field line (defined as $L_f$) between $z = 1$ and the symmetry point can be expressed as elliptic functions, though it is also simple to calculate this length numerically. In the limit of small $y_0$, $L_f \approx 1 - 5\sqrt{y_0}/6-y_0^2/6$. We can also define the cross-sectional area of a flux element ($A(s)$) associated with a given field line by assuming $A(s) \sim 1/B(s)$ where $s$ is a field-aligned coordinate\footnote{Obviously not though at the seperatrix.}. The base area ($A_b$, evaluated at $z = 1$) is given by $A_b = 1/\sqrt{1 + y_0^2}$ and the area at the symmetry line by $A_s \approx 1/\sqrt{2y_0}$. The average area between base and symmetry line will be an important quantity in our analysis and is defined as $\langle A \rangle = (1/L_f)\int_{0}^{L_f} A(s)ds$. For small $y_0$, $\langle A \rangle \approx {\rm{ln}}(\sqrt{y_0})/L_f$. 

\citet{johnson_2024} presented analytic solutions for the temperature profile in the null field for two different heating functions. 
For uniform heating (i.e. $f(B) = 1$) they found:
\begin{eqnarray}
G &=& G_b - \frac{H^*}{4} \Bigl [ (1 - y^2)(1 - z^2) + (1+y^2)(1-z^2)\ln |y| \nonumber \\
 & & + (1+z^2)(1-y^2)\ln |z| \ \Bigr ]\; ,
\label{eq:sol1}
\end{eqnarray}
where $G_b$ is evaluated at $y$ or $z = 1$. These boundaries are assumed to have a lower temperature than the rest of the domain so that a conductive flux passes outward through them. Note also the singularity near the null and along the seperatrices ($y$ or $z = 0$). The right hand panel of \autoref{fig:null_sketch} shows a contour plot of the temperature with a maximum value adjacent to the null and the seperatrices of 10 MK.\footnote{The smallest values of $|y|$ and $|z|$ used in this plot are 0.002, so that the singularities in the solution are avoided. This plot is equivalent to Figure 2 of \citet{johnson_2024}, though with higher temperatures.} The maximum value of G and so T lies along the line of symmetry points $y = z$ and is given by:
\begin{eqnarray} \label{eq:sol2}
G &=& G_{s} = G_b - \frac{H^*}{4} \Bigl [ (1 - y^2)^2 + 2(1-y^4)\ln |y| \Bigr ] \; .
\end{eqnarray}

This solution can be extended to include a constant guide field in the $x$-direction of magnitude $B_p$ so that: 
\begin{equation}
\label{eq:B2}
{\bf B} = \left ( B_p, y, -z \right ) \; ,
\end{equation}
where the $B_0/L$ factor is again suppressed. $G$ then satisfies:

\begin{equation}
\begin{split}
G = G_b(A_x) + \frac{H^*}{2} \Bigl \{\sqrt{\xi^2 + A_x^2} - \xi \sinh^{-1}\left (\frac{\xi}{A_x}\right ) \\
- \frac{B_p^2}{4}\left [ \sinh^{-1} \left (\frac{\xi}{A_x}\right )\right ]^2\ \Bigr \} \; .
\label{eq:gf1}
\end{split}
\end{equation}
Here $\xi = (y^2 - z^2)/2$ where $\xi$ is related to the distance along a field line \citep{johnson_2024}. $G_b(A_x)$ is again chosen to satisfy the boundary conditions that $G = G_b$ at $y = \pm 1$ for all $z$ and $z=\pm 1$ for all $y$. This leads to:
\begin{equation}
\label{eq:gf2}
\begin{split}
G_b(A_x) = G_b - \frac{H^*}{4} \Bigl \{ (1 + A_x^2) \\ - (1- A_x^2)\sinh^{-1}\left (\frac{1- A_x^2}{2 A_x}\right )  \\ - \frac{B_p^2}{2}\left [ \sinh^{-1} \left (\frac{1 - A_x^2}{2 A_x}\right )\right ]^2\ \Bigr \} \; .
\end{split}
\end{equation}
As $A_x\rightarrow 0$, $\sinh^{-1}$ tends towards a logarithm and the $B_p^2$ term will tend towards the square of a logarithm so that the singularity
is more severe in this case than when $B_p = 0$. This is due to the infinite field line length along the seperatrices when $B_p \ne 0$. Note that there are no boundary conditions in the $x$ direction, as we assume x-invariance.

Along the symmetry line, we have
\begin{equation}
\begin{split}
G =  G_{s} = G_b + \frac{H^*}{4} \Bigl \{ - (1 - A_x)^2  \\ + (1- A_x^2) \sinh^{-1}\left (\frac{1- A_x^2}{2 A_x}\right ) \\ + \frac{B_p^2}{2}\left [ \sinh^{-1} \left (\frac{1 - A_x^2}{2 A_x}\right )\right ]^2\ \Bigr \} \; ,
\end{split}
\label{eq:gf3}
\end{equation}
with $A_x = y^2$ there. For fixed $A_x$, the value of $G$ rises, as $B_p$ is increased, so that the temperature rises as the field lines become longer. 

\citet{johnson_2024} also obtained a solution when $f(B) = B^2 = y^2 + z^2$:
\begin{equation}\
\label{eq:hB2}
G = G_b +  \frac{H^*}{8}(1- y^4)(1- z^4)\; .
\end{equation}
This gives solutions that have a flat temperature profile away from the boundaries and no singularities \citep[see][Figure 2]{johnson_2024}.
Finally, for completeness, the solution for $B_p \neq 0$ for this heating function is presented in the Appendix. We do not discuss it further.

\subsection {Conductive cooling at a null}

Conductive cooling in a static atmosphere (no mass motions) is determined by:
\begin{equation}
\frac{1}{\gamma-1} \frac{\partial p}{\partial t} = {\bf B}\cdot \nabla \left( \frac{\kappa_0 T^{5/2}}{B^2} {\bf B} \cdot \nabla T\right)  \; ,
\label {eq:cond1}
\end{equation}
so that $p = p(t)$, $n = n({\bf r})$, $T({\bf r},t) = p(t)/(2k_Bn(\bf{r}))$. If we redefine $G = (p_0/2k_Bn)^{7/2}$, where $p_0 = p(t = 0)$, one finds that, on setting $\gamma = 5/3$:
\begin{equation}
\frac{21L^2}{4\kappa_0} \Bigl (\frac{p_0}{p} \Bigr ) ^{7/2} \frac{dp}{dt} = {\bf B}\cdot \nabla \left( \frac{1}{B^2} {\bf B} \cdot \nabla G\right)  
\label {eq:as1}
\end{equation}
and the left and right hand sides are functions of only time and space respectively\footnote{Following AS76, we solve here for the density rather than the temperature}. As was first noted by AS76, this permits separable solutions, with a separation constant $-k^2$. The left hand side integrates to give:
\begin{equation}
    p(t)/p_0 = 1/(1 + t/\tau_{cool})^{2/5} \; ,
    \label{eq:as2}
\end{equation}
where 
\begin{equation}
    \tau_{cool} = \frac{21L^2p_0}{10k^2\kappa_0}
    \label{eq:as3}
\end{equation}
and $L$ is the spatial normalisation, {\it not} the length of an individual field line ($L_f$). The right hand side is then:
\begin{equation}
{\bf B}\cdot \nabla \left( \frac{1}{B^2} {\bf B} \cdot \nabla G\right)  = -k^2 \; .
\label{eq:as4}
\end{equation}
The time dependence of the temperature is thus determined by $\tau_{cool}$ which in turn depends on the value of $k^2$.  

\subsection{Methods of solving the conduction equation}

There are two ways to determine $k^2$.
For constant heating, the equation for the equilibrium between heating and cooling \eqref{eq:en2}, and for the spatial temperature structure during cooling \eqref{eq:as4}, are of an identical form with the substitution of $H^*$ for $k^2$. Thus, one can set $k^2 = H^*$ in the definition of $\tau_{cool}$. This is a powerful general result that applies to all situations when the heating is initially spatially constant and then turned off abruptly. It means that the cooling equation does not have to be solved. 

Alternatively, the field-aligned version of \eqref{eq:as4} is solved subject to a symmetry condition ($dG/ds = 0$) at the symmetry point of each field line and a boundary condition on $G$ along $z = 1$ such that $G = G_b$ there, with $G_b$ assumed "small", i.e. $G_b \ll G_s$. For a given heating, the quantity $G = G_s$, and hence the temperature at the symmetry point $T = T_s$, is determined from \eqref{eq:sol2}, this in turn determines $k^2$. Numerically, the solution of \eqref{eq:as4} subject to these boundary conditions requires a double iteration involving $k^2$ and the base heat flux. [For a straight field, $k^2 = 2 G_s(L^2/L_f^2)$ where $G_s \gg G_b$, so that $\tau_{cool} = 21L_f^2p_0/(20\kappa_0T_s^{7/2})$ as found by AS76.] 

\subsubsection{Pressure and density models for null point cooling}

We have not yet discussed the behaviour of the pressure and density in the null point model. They are of importance since they directly impact conductive cooling (see \eqref{eq:as3}) but do not arise in the expressions for the equilibria. Both the equilibria and cooling models have no mass flow, so that the pressure is constant along field lines and $n(s) \propto 1/T(s)$ . Further, since there is no Lorentz force exerted by the magnetic field, the pressure should be the same for all values of $y$ and $z$ which implies that the density at the symmetry points should scale as $1/T_s$. For constant heating, $\tau_{cool}$ is then the same for all field lines since $k^2 = H^*$. This result also holds for any specified value of $B_p \neq 0$. 

An alternative approach treats the density as the same at all the symmetry points, implying a pressure gradient across the field lines. So long as the plasma pressure is small compared to the magnetic pressure, one can adopt this approach and still treat each field line as a mini-atmosphere. However, this is not necessarily true around the null, and we present an extensive discussion of the implications of this in the Conclusions.

\section{Cooling results}
\subsection{Constant heating, no guide field}
First we consider constant heating with no guide field. Although the case with $B_p = 0$ is a subset of the more general solutions, we treat it separately in order to highlight aspects of our approach. In line with the motivation discussed in the Introduction, the initial conditions have (maximum) temperatures of 5 - 10 MK at the symmetry point of the field lines. For a system box size with L = 25 Mm, this requires $H_0 = 0.05$ with $H^*$ following from its earlier definition. $y_0$ is taken as the fundamental parameter in what follows and we solve for 29 values between $10^{-5}$ and 0.85 which are logarithmically equally spaced according to the formula $y_0(j) = 10^{-5}\times 1.5^{j-1}, ~ j=1,29$. This gives good coverage over a range of $y_0$ which is important for benchmarking the analytic solutions presented later. 

As noted in Section 2.2.1, two cases of the plasma properties at the symmetry points are considered. For the pressure assumed to be constant everywhere, we set $n = 10^9 \rm cm^{-3}$ at the symmetry point with the smallest value of $y_0 (= 10^{-5})$ which in turn gives a pressure of 2.69 dyne $\rm cm^{-2}$. Since the pressure is the same at all the symmetry points, and the temperatures there decrease as $y_0$ increases, the density at the symmetry point must correspondingly increase from $10^9 \rm cm^{-3}$ to $5.83 \times 10^9 \rm cm^{-3}$. On the other hand, if the density is the same at all the symmetry points ($10^9 \rm cm^{-3}$), the pressures there decrease from 2.69 to 0.46 dyne $\rm cm^{-2}$ as $y_0$ increases.

\begin{figure*}

  \includegraphics[width=0.45\linewidth]{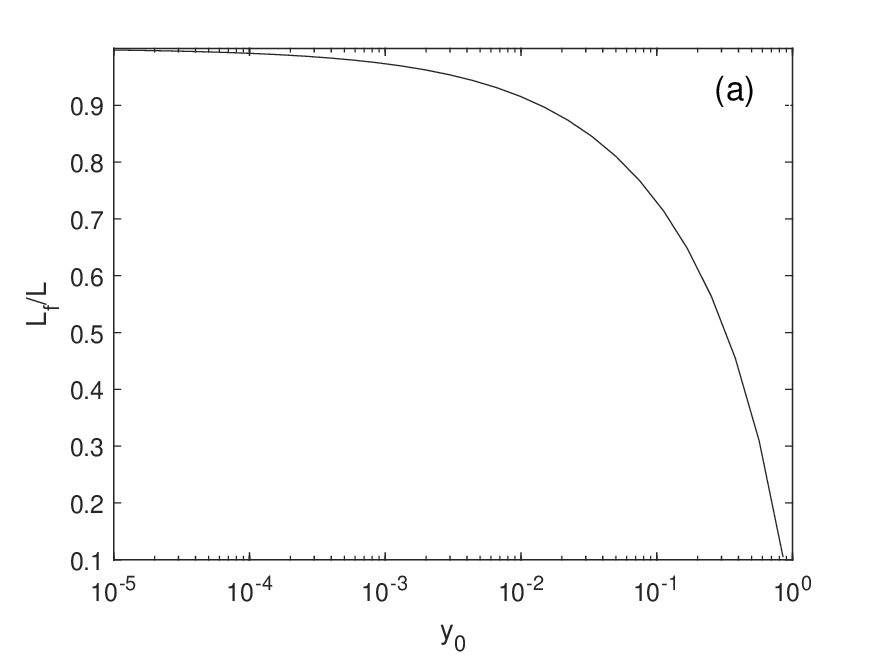}
  \includegraphics[width=0.45\linewidth]{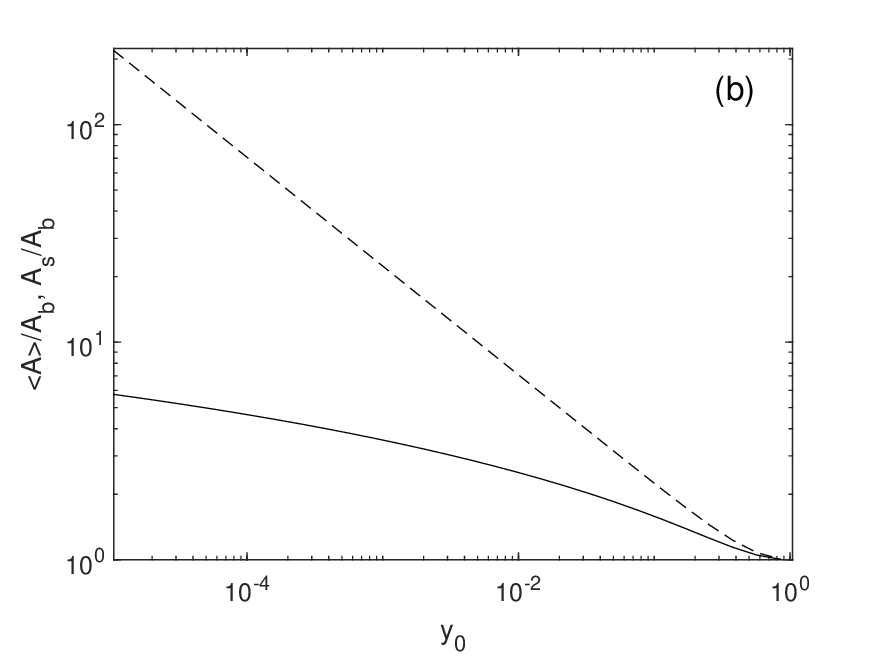}

  \includegraphics[width=0.45\linewidth]{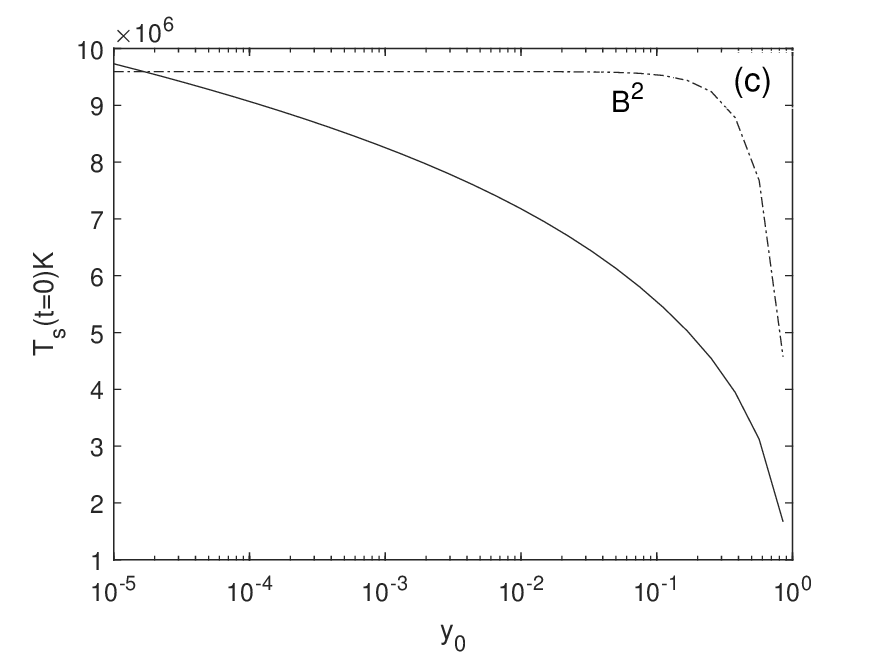}
 \includegraphics[width=0.45\linewidth]{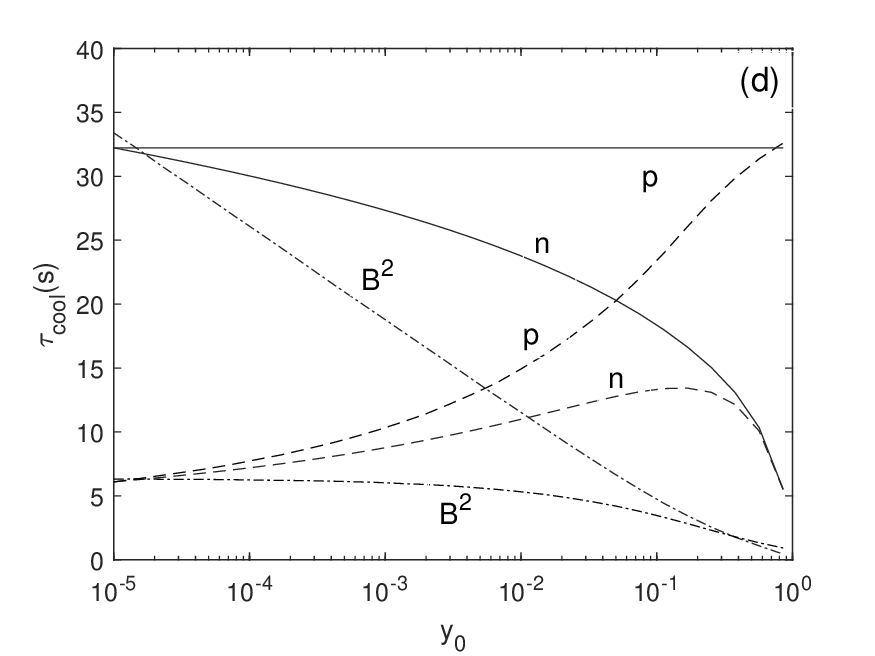}
   
	\caption{Results for the cooling of equilibria \eqref{eq:sol1}. All plots are as a function of $y_0$. Upper left and right (a and b respectively) show the length of a normalised field line ($L_f/L$) and the apex (dashed) and average (solid) area respectively, normalised to the area at $z = 1$. Lower left (c) shows the initial temperature ($T_s$) at the symmetry point of each field line. Lower right (d) shows $\tau_{cool}$, with the black solid curves labelled "p" and  "n" corresponding to the constant pressure and density cases (see text). The dash-dot curves labelled $B^2$ correspond to equilibrium \eqref{eq:hB2}. The three curves beginning at $\tau_{cool} = 6$ correspond to respective cases (as labelled) with a uniform area, but the same initial temperature and field line length.}
	\label{fig:noguide}
\end{figure*}

In \autoref{fig:noguide} the top two plots (labelled a and b) show the ratio of the field line length ($L_f$: defined earlier) to the box size, and the average (solid curve) and apex (dashed curve) areas normalised to the area at $z = 1$, all as a function of $y_0$. Both areas decrease as $y_0$ increases: the apex area for small values of $y_0$ is large, in excess of 100. However, the area averaged over the field line shows much smaller variation. [Note that neither $L_f$ nor $\langle A \rangle$ are needed to calculate $\tau_{cool}$ here. The remarkable thing about the $H^* \rightarrow k^2$ substitution is that these effects are all built in. However, as we will show shortly, knowing $L_f$ and $\langle A \rangle$ are essential for a physical understanding of the cooling process.] 

The lower left panel (c) shows the initial temperature at the symmetry line, which decreases as $y_0$ increases. The lower right panel (d) shows $\tau_{cool}$ which for the present case of constant heating has been obtained both by the substitution $H^* \rightarrow k^2$ and the numerical solution of the boundary value problem \eqref{eq:as4}. For constant pressure, the $H^* \rightarrow k^2$ substitution method gives $\tau_{cool} = 32.2$ sec for all $y_0$, and this is shown by the upper straight line labelled \lq p\rq\  in panel d. The values of $\tau_{cool}$ obtained from the boundary value problem agree with the above number to within 0.01\% for the first 27 values of $y_0$, increasing to 0.5 and 1\% for the last two for our convergence criteria. This confirms the validity of the $H^* \rightarrow k^2$ substitution.

The values of $\tau_{cool}$ for the case of constant density at the symmetry points are shown by the solid curve labelled \lq n\rq\  in the same panel. We now see that $\tau_{cool}$ 
decreases as $y_0$ increases which is due to the decrease in the pressure at the symmetry points (see \eqref{eq:as3}). Although not strictly comparable since the pressure along the symmetry line varies, the substitution method and numerical solution show the same level of agreement as before.

In both these cases the behaviour of $\tau_{cool}$ is the result of an interplay between the initial maximum temperature and pressure, the area profile along the field, and the field line length. To understand this, we determine an approximate solution to the static, time-dependent conduction equation with variable area \citep[see for example][] {cargill_2022}. Integrating once along the field line gives:
\begin{equation} \label{eq:ebtel1}
    \frac{\langle A \rangle L_f}{\gamma -1} \frac{dp}{dt} = A_b F_b 
\end{equation}
 where $F_b$ is the heat flux at the base. Note that given our previous approximations, this is an exact result. 
 Following the Enthalpy-Based Thermal Evolution of Loops (EBTEL) approach \citep{ebtel_2008,cargill_2022}, we approximate\footnote{The choice of the constant 4/7 here as opposed to the usual 2/7 is discussed in a subsequent paper. We simply note that the AS76 solution corresponds to the value of 4/7 used here.} $F_b \approx - (4/7) \kappa_0 T_s^{7/2}/L_f$. This equation can be integrated to give:
 \begin{equation} \label{eq:ebtel2}
     T_s/T_s(t=0) = 1/(1 + t / \tau_*)^{2/5} \; ,
 \end{equation}
 where
 \begin{equation} \label{eq:ebtel3}
    \tau_* =  
    \frac{21k_Bn_sL_f^2}{10\kappa_0 T_s^{5/2}} \frac{\langle A \rangle}{A_b}
    = \frac{21p_0L_f^2}{20\kappa_0 T_s^{7/2}}\frac{\langle A \rangle}{A_b}
\end{equation}
with $p_0$, $n_s$ and $T_s$ evaluated at the start of the cooling phase. This is the AS76 result modified by an area factor. Note also the superficial similarity to \eqref{eq:as2} but now the separation constant is no longer present and the actual field length appears. Substituting the values of $T_s$, $L_f$ and $\langle A \rangle$ from \autoref{fig:noguide} into these expressions, we find that for both constant pressure and constant density cases, $\tau_*$ agrees with the values of $\tau_{cool}$ shown in \autoref{fig:noguide} to between 10 and 15  \% with $\tau_{cool} < \tau_*$. While the scaling of $\tau_{cool}$ with length, temperature and density are familiar, a new aspect is that we are able to approximate the rapidly varying area around the null simply by its average along the field line. This holds even in the extreme cases where the area near the null is very large\footnote{Note though that the scaling $\tau_{cool} \sim \langle A \rangle$ is not a general one for all area profiles, as we will discuss shortly.}. Thus, it is $\langle A \rangle$ not $A_s$ that determines the cooling properties.

To further quantify the effect of non-uniform area on cooling, it is instructive to compare the above results to those with the same initial temperature, pressure and field line length, but uniform area. We do this in an approximate way by taking the initial state (i.e. $T_s$ and $L_f$) and solving the cooling equation for $\langle A \rangle = A_b$ to then determine $k^2$. These are shown by the dashed lines labelled \lq p\rq\  and \lq n\rq\  in panel d of \autoref{fig:noguide}. The cooling for the constant pressure case is now no longer the same for all field lines. It is also seen that for both cases as $y_0$ approaches unity, the respective values of $\tau_{cool}$ approach each other since the null point field tends towards being uniform for large $y_0$. The difference between the null and constant field values of $\tau_{cool}$ scales as $\langle A \rangle$ as expected from the above discussion.

\subsection{Constant heating, guide field included}

\begin{figure*}

  \includegraphics[width=0.45\linewidth]{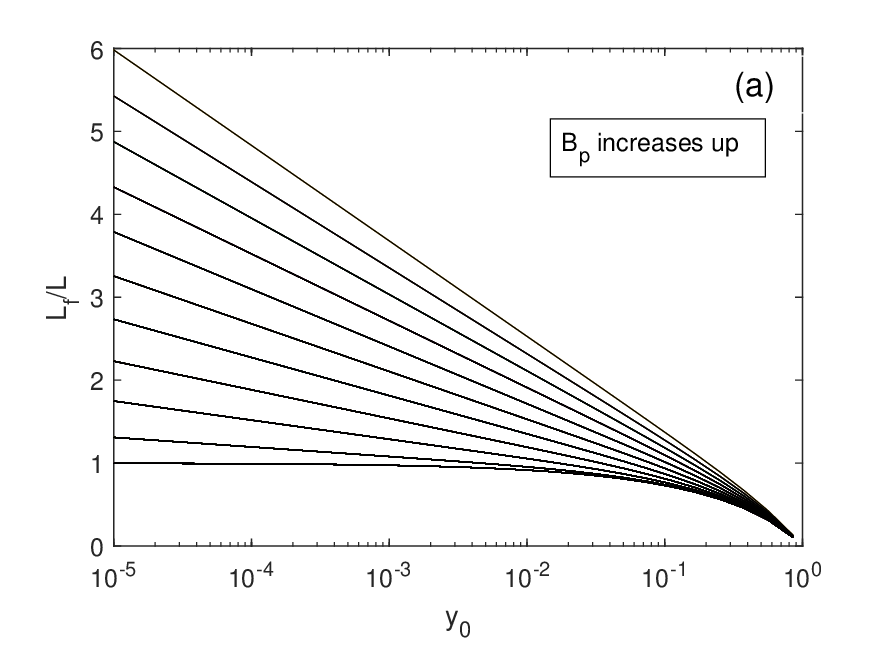}
  \includegraphics[width=0.45\linewidth]{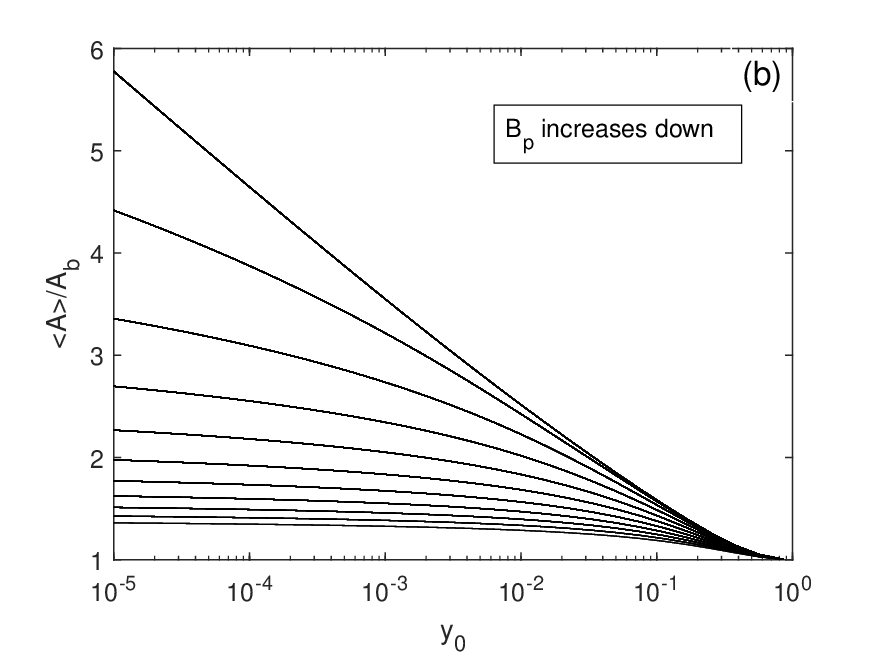}

     \includegraphics[width=0.45\linewidth]{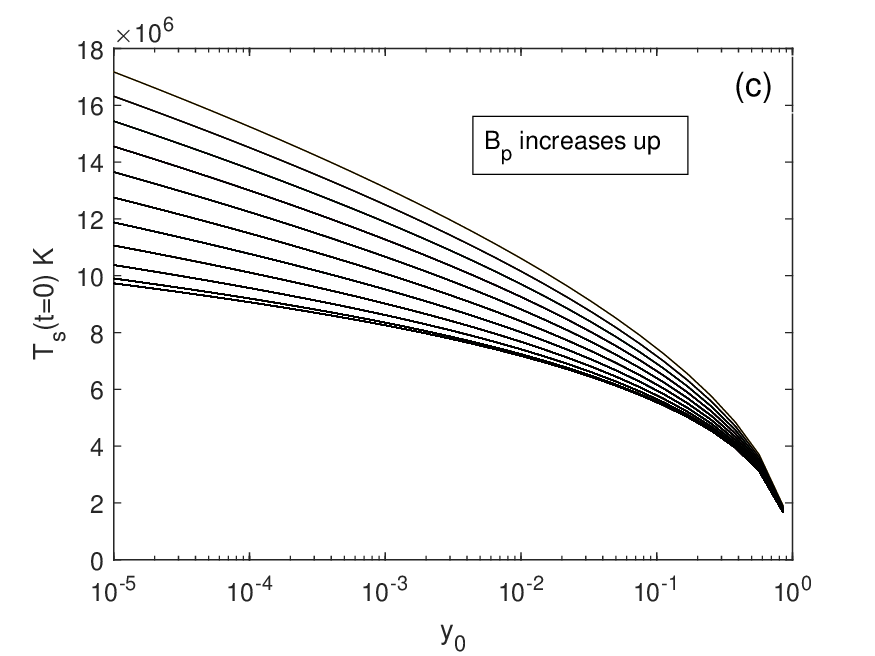}
  \includegraphics[width=0.45\linewidth]{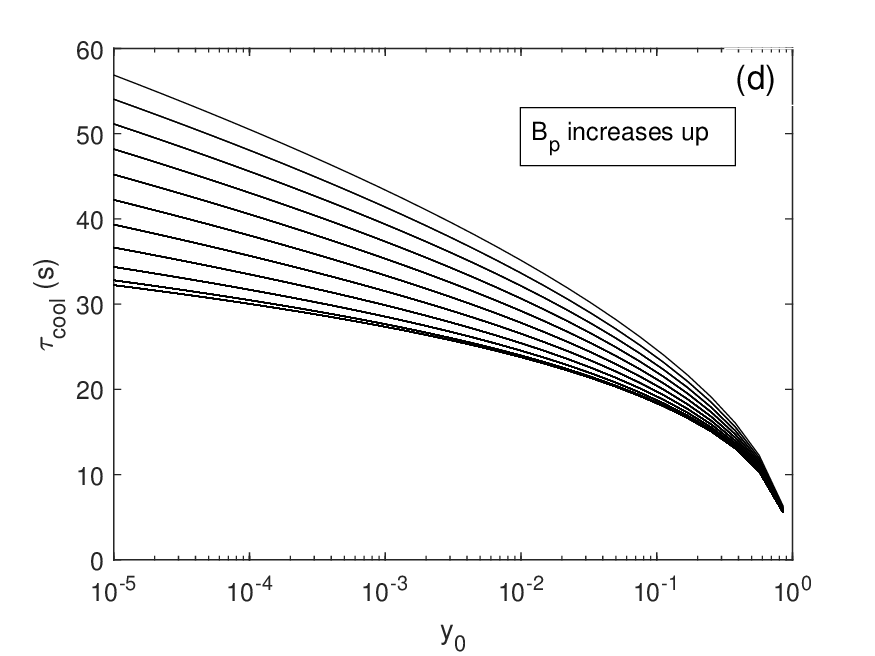}
  
	\caption{As \autoref{fig:noguide} except a guide field is now included. $B_p$ increases in equal steps from 0 to 1 as shown by the captions in each panel. The values of $\tau_{cool}$ are for constant density at the symmetry points. For constant pressure, $\tau_{cool}$ is a straight horizontal line from the smallest $y_0$.}
	\label{fig:guide}
\end{figure*}

When the guide field is included, the same $H^* \rightarrow k^2$ substitution in the expression for $\tau_{cool}$ can be made. \autoref{fig:guide} shows the results in the same format as \autoref{fig:noguide}. All panels show 11 values of $B_p$ uniformly distributed between 0 and 1, increasing or decreasing from lower to upper curves as indicated on each panel. The upper two panels demonstrate that as $B_p$ increases, the field line length increases while $\langle A \rangle$ decreases, the latter result arising because the constant guide field becomes increasingly dominant\footnote{Note that the comparison with the straight field case thus becomes less relevant as $B_p$ increases.}. Fixing $H^*$ to the same value as before, the lower left panel shows that the initial temperature increases with $B_p$, as can be seen in \eqref{eq:gf3} and is due to the longer field lines\footnote{It is interesting to note that while $B_p$ of order unity introduces a significant change in $\tau_{cool}$, it corresponds to a modest shear in the field. The field in the $y-z$ plane at the boundaries is $\sqrt{1 + y_0^2}$ so for $B_p = 1$ the shear there is less than $45^o$.}. 

For each $B_p$ a different set of values of $\tau_{cool}$ are obtained (lower right panel). For the case of constant density at the symmetry points and a given $y_0$, increasing $B_p$ leads to larger values of $\tau_{cool}$. The constant pressure values of $\tau_{cool}$ are straight horizontal lines for each value of $B_p$, for example for $B_p = 1$, $\tau_{cool} = 56.9$ sec for all values of $y_0$. As before, solving the differential equation for $k^2$ gives the same results as the $H^* \rightarrow k^2$ substitution $H^*$. 

Once again, these results arise due to the interplay of $\langle A \rangle$, $L$ and $T_{max}$. For both cases, the values of $\tau_{cool}$, as $B_p$ increases, arise from the competition between higher initial temperature and smaller area changes (both making cooling more effective), and longer field lines (slows cooling). For constant density, the latter wins out. For constant pressure, as before, these effects compensate each other exactly. 

The two limits of very small and large $B_p$ correspond to the case discussed in Section 3.1 and a uniform area respectively and show good agreement between $\tau_{cool}$ and $\tau_*$. Intermediate values of $B_p$ do not give such good agreement between $\tau_*$ and $\tau_{cool}$. As an example, for $B_p = 0.3$, the discrepancy is of order 60\% for $y_0 = 10^{-5}$, decreasing to 25\% for $y_0 = 10^{-2}$ and showing good agreement as $y_0$ exceeds 0.1. In all cases $\tau_{cool} < \tau_*$. The difference between $\tau_{cool}$ and $\tau_*$ lies in the approximation for $F_b$ used in calculation of $\tau_*$ (see above Eq(17)), where the factor 4/7 is too small. The details of this are the subject of a detailed study in a subsequent paper, but in general $F_b$ is determined by the temperature profile over the entire spatial domain. When the area variation is predominately near the symmetry points, or the area is uniform, the factor $4/7$ is adequate. When the area variation is less concentrated near the symmetry points, the change in the temperature structure along the entire field line then gives a larger value of $F_b$ than is modelled by the 4/7 factor.


\subsection{Cooling of an approximately isothermal source}

In this case there is no equivalence between the heating function and the separation constant $k^2$ and for the cases we have looked at, $k^2$ shows considerable variation with $y_0$. In turn, this means that the spatial temperature structure differs between the equilibrium and cooling, a result verified by a direct comparison of the two solutions. This change in temperature profile implies a change in density and corresponding mass motions. The neglect of these requires that the characteristic conduction time is significantly smaller than the sound travel time. We have compared $\tau_{cool}$ with $\tau_{sound} = L_f/C_s$ where $C_s = \sqrt{2k_BT_s/m_p}$. For all but the smallest $y_0$ we find $\tau_{cool} \ll \tau_{sound}$. What will happen is that, when the heating is turned off, the initial equilibrium will adjust rapidly by thermal conduction to the new profile quickly enough that the density does not have time to respond. For very small $y_0$, the density can respond weakly. This conclusion has been confirmed by simulations with the Lare code (see below).

\begin{figure*}

  \includegraphics[width=0.45\linewidth]{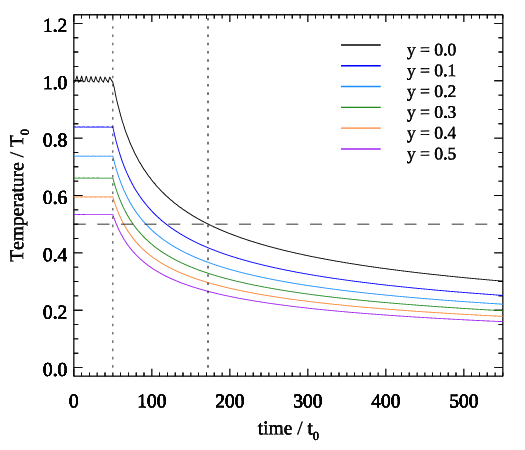}
  \includegraphics[width=0.45\linewidth]{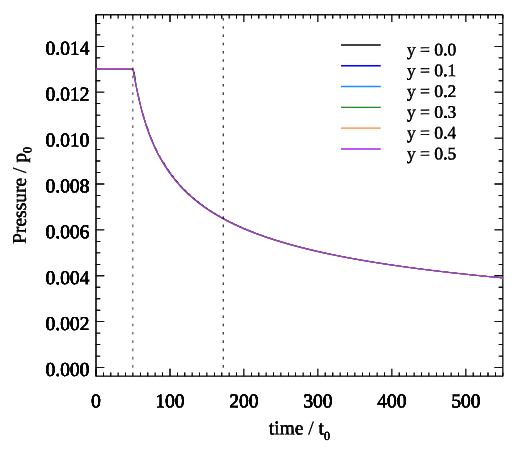}

     \includegraphics[width=0.45\linewidth]{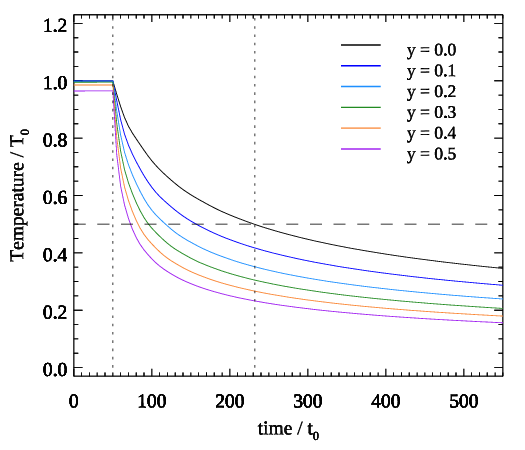}
  \includegraphics[width=0.45\linewidth]{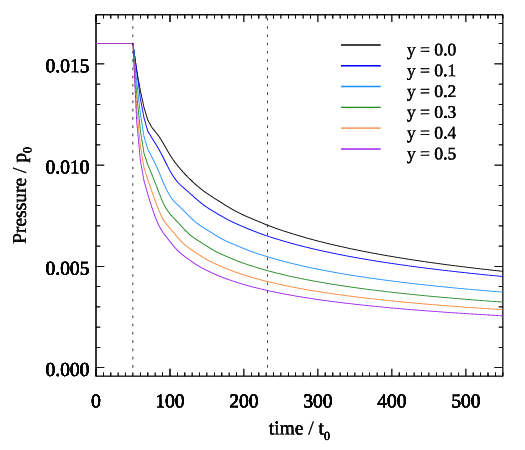}
   
	\caption{Lare results for the cooling of equilibrium 5 (constant heating: upper panels) and 10 ($B^2$ heating: lower panels). The left (right) hand plot shows the evolution of the temperature (pressure) at 6 locations along the symmetry line. For constant heating the pressure results lie on top of each other. The dashed and dotted lines are discussed fully in the associated text. In all cases Lare dimensionless units are used.}
	\label{fig:lare}
    \end{figure*}

$H_0$ is now increased to be 1, so that the maximum temperature at the symmetry points is still of order 10 MK.
The upper dash-dot lines labelled $B^2$ in panels c and d of \autoref{fig:noguide} show the initial temperature and $\tau_{cool}$ for equilibrium \eqref{eq:hB2} with the apex density constant. As noted above, the temperature is roughly constant along $y = z$ until very near the corner $y_0 = 1$. Compared to the constant heating case, $\tau_{cool}$ decreases more rapidly as $y_0$ increases. Since $L_f$ and $\langle A \rangle/A_b$ are the same for both cases, this can be attributed to the higher initial temperatures at the symmetry points. The very short $\tau_{cool}$ near $y_0 = 1$ (in fact $y_0 = 0.852$) compared with the constant heating case is because the initial temperature remains very high there. $\tau_{cool}$ again agrees well with the approximate scaling described above. Finally the lower dash-dot line shows $\tau_{cool}$ for an imposed uniform area: as before it is reduced by an approximate factor $A_b/\langle A \rangle$.

\subsection{MHD simulation results}
    
As a further verification of these results, we have performed simulations of null point cooling using a 2.5D version of the MHD Lare code \citep{arber_2001,reid_2020} which includes field-aligned and cross-field conduction \citep[see][]{johnson_2024}, as well as plasma flows and time-dependent magnetic fields, though last three of these only produce small effects.
\autoref{fig:lare} shows the evolution of the temperature (left) and pressure (right) in Lare dimensionless units for six different field lines, defined by $y = \epsilon...... 0.5$ where $\epsilon \ll 1$ is a distance of a half width of grid cell from the null. The initial state is shown on the left of each panel with the field lines labelled in the captions, and each field line having a different initial temperature. The upper and lower rows show the results for constant heating and heating proportional to $B^2$ respectively. The initial temperatures are given by Eq (5) and Eq (10), and a specified spatially constant pressure (see Section 2.1.2), with the density then derived from the equation of state.

At t = 50 (first vertical dotted line in all panels), the heating is abruptly turned off and the plasma allowed to cool. For the constant heating case, the pressure shows an identical decay on all field lines : the six curves lie on top of one another. This is not the case when the heating is proportional to $B^2$. Next, we look at the time to cool by a factor two (see Eq 19). In the left hand panels, the horizontal dashed line indicates the temperature at that time associated with the hottest initial condition, and the second dotted vertical line shows the time it takes to reach that temperature, approximately 120 time units. 
For constant heating, the five other initial temperatures also take 120 time units to fall by a factor of two. Thus, the simple $H^* \rightarrow  k^2$ substitution proposed earlier for constant heating is confirmed with these simulations, while for heating proportional to $B^2$ this does not arise.

\section{Discussion}

The results in the previous section have been presented in terms of the timescale $\tau_{cool}$ and this is a very convenient way to understand the scalings of conductive cooling on the geometrical and plasma parameters at the null. However, we have deliberately not referred to this as a \lq cooling time\rq \  since that quantity depends on the temperature range one wishes to study the cooling over which, in turn, requires the use of \eqref{eq:as2}. 

\begin{figure}

  \includegraphics[width=\linewidth]{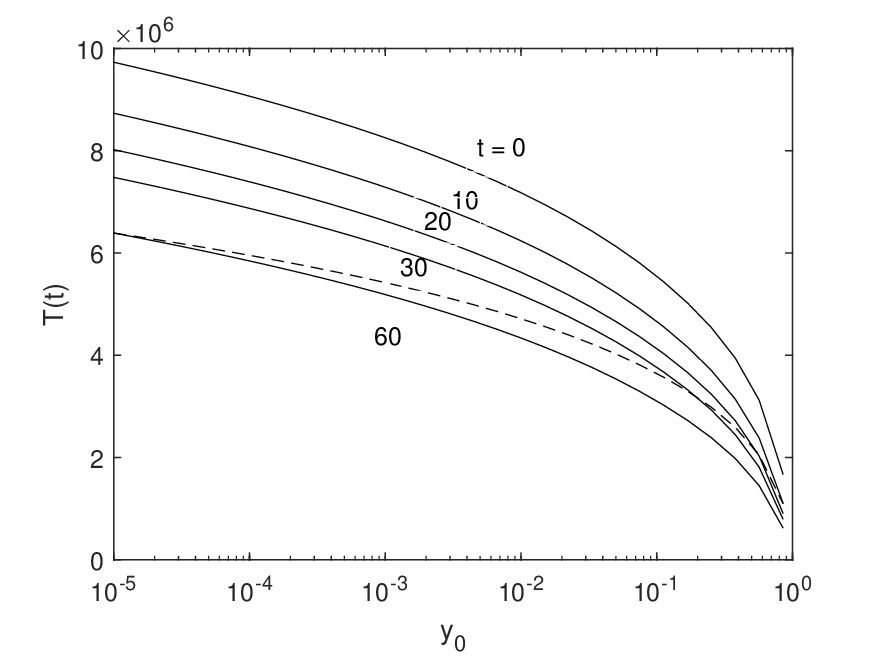}

	\caption{Cooling of the two-dimensional equilibria. The five curves show the temperature at the symmetry points as a function of $y_0$ at the initial time and four subsequent times, as indicated. The dashed curve is the constant pressure result at 60 sec.} 
	\label{fig:cooling_vs_time}
\end{figure}

As an example, \autoref{fig:cooling_vs_time} shows the temperature at the symmetry points as a function of $y_0$ at several different times (t = 0, 10, 20, 30 and 60 secs, as indicated). We see that despite $\tau_{cool}$ ranging from between 30 and a few sec, the time taken to cool by, for example, a factor of two is much longer, in excess of 60 sec. More generally, the time taken to cool from an initial value $T_0$ at $t = 0$ to a new value $T_1$ at $t = t_1$ is:
\begin{equation} \label{eq:tc2}
    t_1 = \tau_{cool} \Bigl [ (T_0/T_1)^{5/2} - 1 \Bigr ] \; ,
\end{equation}
so that cooling by, for example, a factor 2 takes $4.66 \tau_{cool}$. As another example, to cool from 8 MK to 6 MK with an initial state of 10 MK takes 1.85 $\tau_{cool}$. These scalings indicate that the real difference between the cooling times in the null and uniform fields are larger than a simple comparison of the values of $\tau_{cool}$.  Care is thus needed by what is meant by the term \lq cooling time\rq .

This needs to be taken into consideration if we evaluate the actual volume of plasma in a given temperature range around the null. To address this, we first calculate the total `volume' above a specified temperature as a function of time (the volume is actually an area 1 cm thick). To obtain the `volume' associated with each value of $y_0$, we define the width of the flux element associated with a given field line as having a width of half the width to the neighbouring field lines, or in the case of the smallest $y_0$, the distance to the seperatrix. This gives a volume of $3.125 \times 10^{18}$ $cm^2$ between the axis and symmetry line which must then be multiplied by some distance along the x-axis to get a real volume. Taking the latter as 25 Mm gives a real volume of this segment of $7.8 \times 10^{27}$ $cm^3$. The total volume surrounding the null is then 8 times this.

\begin{figure*}

  \includegraphics[width=0.45\linewidth]{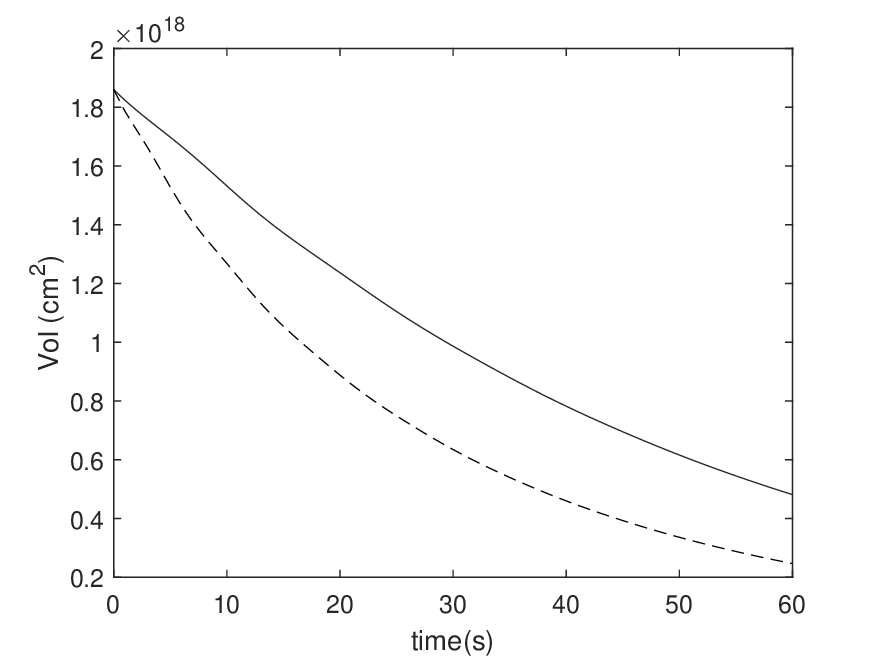}
  \includegraphics[width=0.45\linewidth]{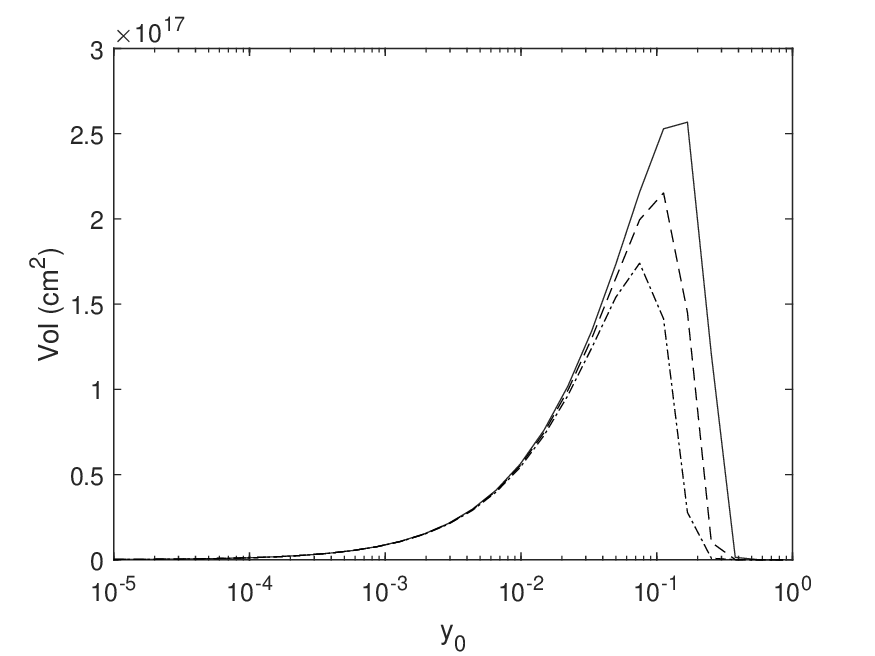}

     \includegraphics[width=0.45\linewidth]{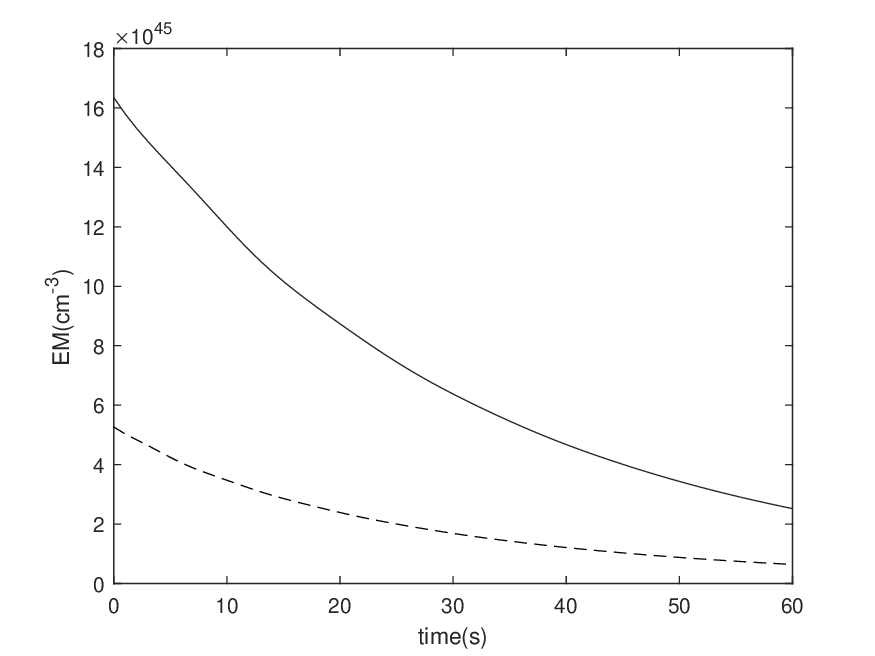}
  \includegraphics[width=0.45\linewidth]{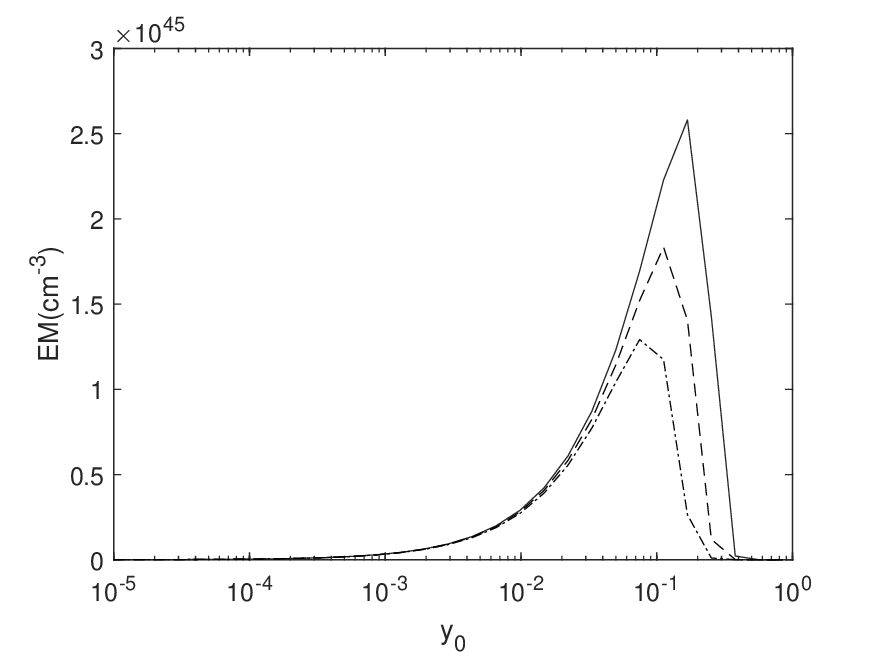}
  
	\caption{Upper left: the volume of hot plasma above 4 MK for equilibria \eqref{eq:sol1} as a function of time with constant pressure (solid) and constant density (dashed) along the symmetry line. The vertical axis needs to be multiplied by a characteristic width in the x-direction to obtain the actual volume (see text). The upper right panel shows the distribution of the hot plasma as a function of $y_0$ for three different times: 10, 20 and 30 sec (solid, dashed, dot-dashed) for the constant pressure case. The lower panels show the emission measures corresponding to the two upper plots with a transverse distance of 25 Mm.}
	\label{fig:hot_vol}
\end{figure*}

The upper left panel of \autoref{fig:hot_vol} shows the hot volume for the two-dimensional case for constant pressure (solid) and constant density (dashed) above 4 MK. \footnote{Note that the almost isothermal nature of the temperature profile, and the limited sampling of field lines, means that such an abrupt temperature cut-off leads to small bumps in the calculated hot volume when any flux element drops below the critical temperature. To mitigate this, we smooth the way the volume is counted through the cut-off temperature using a tanh function centered on the cut-off temperature with a scale of 0.25 MK.} The time evolution of the two heating models is similar with the constant pressure volume somewhat larger since it cools more slowly.

It is also of interest to determine which locations in the field around the null contribute to the hot volume. The upper right panel of \autoref{fig:hot_vol} shows this as a function of $y_0$ at three different times: 10, 20 and 30 sec. The main contribution is located away from the null but moves towards the null as the cooling proceeds. Higher cut-off temperatures lead to smaller volumes localised closer to the null.

The two lower panels show the corresponding emission measures. This takes into account both the behaviour of the density along the symmetry line and along the field lines. We see that the constant pressure case has the larger emission measure due to the need to increase the density along the symmetry line. 

\section{Conclusions}
In this simple examination of null point cooling, we have demonstrated that the non-uniform area profile of the field near the null introduces a slower cooling 
than would arise in a straight field. While the area of a flux element near the null can be one to two orders of magnitude larger than at the boundary, 
it is the average area along the field line that dictates the cooling time. Thus the difference from a straight field is reduced to a maximum factor less than 5 for the two-dimensional null considered here. 
The introduction of a `guide field' leads to a cooling time closer to that of a uniform field. The approximate scaling $\tau_{cool} \sim \langle A \rangle$ is the same 
as that proposed by \citet{cargill_2022} for a loop geometry, and we show analytically that it is a good approximation, at least for the 2D null, 
although the result is not entirely general for all area profiles. We will address this in a subsequent paper.

Three further comments can be made. 
We noted in Section 2.2.1 that, depending on whether the pressure or density was constant along the symmetry line, different pressure gradients in the initial plasma conditions could arise. If these forces are large enough, the field may be distorted from its potential configuration around the null which in turn changes $A(s)$ there from the values used.
For equilibrium (9), the pressure is spatially constant everywhere, with a decreasing value as time increases. For equilibrium (10), \citet{johnson_2024} and panel c of \autoref{fig:noguide} show that the temperature (and hence pressure) is approximately constant for $y_0 < 0.1$ so that the constant density and constant pressure cases are almost identical. In all these cases the pressure forces can then be ignored and we can assume a set of rigid atmospheres, as done in Section 3. 

On the other hand, for equilibrium (9), the constant density assumption does lead to a pressure gradient transverse to the field lines at the symmetry line. If we define the plasma beta there as $\beta_s = \beta_{s0}/((B_p/B_0)^2 + 2y_0)$ with $\beta_{s0} = 8 \pi p_s/B_0^2$, for $B_p = 0$ and our default parameters, $\beta_s = 0.25$ when $y_0 = 10^{-2}$, suggesting that pressure forces may have a significant effect on the field near the null point {\footnote{Note that if $B_p = 0$, there is always a value of $y_0$ below which $\beta_{s0}$ becomes very large. However, finite $B_p$ implies a magnetic pressure everywhere along the symmetry line and the guide field becomes dominant when $B_p/B_0 > \sqrt{2y_0}$. For example, for $y_0 = 10^-5$, near the null, $B_p/B_0 > 4.5 \times 10^{-3}$. So the singularity in beta at the null is readily eliminated by a very small $B_p$.}}. 
However, it is the pressure gradient that is important, or the ratio $R = 8 \pi |(dp/dr)/(dB^2/dr)|$, where r is a coordinate along the symmetry line. $dp/dr$ can be obtained from equation (9) and $dB^2/dr$ from equation (3). For $B_p = 0$, we find that for $y_0 = 2 \times 10^{-3}$, $R < 0.1$, so for $y_0$ larger than this a rigid field can be assumed. For small $B_p = 0.02$, the same value of R is obtained. These small limiting values of $y_0$ arise because the plasma pressure falls off with r much more slowly than the magnetic pressure increases.

This suggests that for small $y_0$ a full MHD simulation would be required, though with a resolution of order one to two orders of magnitude better than that used in Section 3.4. This is unfeasible at the present time. We also note that \autoref{fig:hot_vol} shows that the contribution to the emission of the field lines associated with $y_0 < 10^{-2}$  is small. Thus while the scalings derived earlier apply for all values of $y_0$, we caution that there may be small changes in the area profile of the field lines associated with small $y_0$, though this will also depend on parameters such as $B_0$ and the heating function.

Secondly, optically thin radiation has been neglected, as is appropriate for the temperatures and density at the start of the cooling phase. However, it is well known that when a heat flux from the corona interacts with the transition region and chromosphere, a process known as chromospheric evaporation takes place, increasing the coronal density (see \citet{as78} for background and \citet{sb_pc2013} for a more modern viewpoint). This in turn will negate the asssumption of the neglect of radiation, leading to a regime where conduction and radiation are comparable and finally to a purely radiative cooling stage \citep{cargill_1994}. We have elsewhere argued \citep{cargill_2022} that an area factor such as discussed here has a small effect on the temperature decay in the radiative phase but leads to an over-dense corona compared to a field profile with uniform area. Clearly this is beyond our present scope, requiring a realistic model of the corona / transition region interface, but these considerations should provide a guide for future work.

Finally the extension to 3D nulls is also an open question. The geometry \citep[e.g.][]{priestforbes} is more complicated and is not amenable to analytic considerations due to its fully 3D nature, even for the simplest fields such as ${\bf B} = (-2x, y, z)$ in our notation. Of particular difficulty is the transition between the cross-section of a flux element as one goes from close to the spine (y and z small) to the fan plane (x small) and without direct calculations, no rigorous conclusion can be drawn. However, we note that the basic principle of a strongly divergent field near the null can still hold, and one can then expect to see inhibition of cooling of hot sources at such a null.

\section*{acknowledgements}
AWH and DJ gratefully acknowledge the financial support of the Science and Technology Facilities Council (STFC) through Consolidated Grants ST/S000402/1 and ST/W001195/1 to the University of St Andrews.

\section*{Data Availability}
The Lare3D code, used to produce the simulations in this work, is freely available and may be found at https://github.com/Warwick-Plasma/Lare3d and is described in \citet{arber_2001}.

\bibliography{references}{}
\bibliographystyle{mnras} 

\onecolumn
\appendix
\section{$B^2$ heating and a guide field}

The solution for $G$ for heating with a constant component and one proportional to $B^2$ in the presence of a guide field is:
\begin{eqnarray*}
G&=& G_b - \frac{H_0}{4}(1-y^2)(1-z^2) -\frac{H_1}{8}(1 -y^4)(1 - z^4 ) \; \\
&&-\frac{1}{4}((H_1B_p^2 + H_0)(y^2 - z^2)\sinh^{-1}\left (\frac{y^2 - z^2}{2yz}\right ) + \frac{1}{4}(H_1B_p^2 + H_0)(1 -y^2z^2)\sinh^{-1}(\frac{1 -y^2z^2 }{2yz})\; \\
&& - \frac{B_p^2}{8}(H_1B_p^2 + H_0)\left [\sinh^{-1}\left(\frac{y^2 - z^2}{2yz}\right )\right ]^2 + \frac{B_p^2}{8}(H_1B_p^2 + H_0)\left [\sinh^{-1}\left(\frac{ 1 -y^2z^2}{2yz}\right )\right ]^2\; .
\end{eqnarray*}
where $H_0$ and $H_1$ are the constant and $B^2$ heating magnitudes.

\label{lastpage}
\end{document}